\documentstyle[times,balanced,epsfig,pre,aps]{revtex}

\newcommand{\be}{\begin{equation}}
\newcommand{\ee}{\end{equation}}
\newcommand{\bc}{\begin{center}}
\newcommand{\ec}{\end{center}}
\newcommand{\bi}{\begin{itemize}}
\newcommand{\ei}{\end{itemize}}
\newcommand{\ba}{\begin{eqnarray}}
\newcommand{\ea}{\end{eqnarray}}

\def\dirfig{.}

\begin{document}


\title{
The Complex Ginzburg-Landau Equation in the Presence of Walls and
Corners.
}

\author{
V\'\i ctor M. Egu\'\i luz$^{1,2,}$\cite{email}, Emilio Hern\'andez-Garc\'\i
a$^2$ and Oreste Piro$^2$ \\
}
\address{
$^1$Center for Chaos and Turbulence Studies\cite{cats}, The Niels Bohr Institute,
Blegdamsvej 17, DK2100 Copenhagen \O (Denmark) \\
$^2$Instituto Mediterr\'aneo de Estudios Avanzados
IMEDEA\cite{imedea}(CSIC-UIB),
E-07071 Palma de Mallorca (Spain)
}

\date{\today}

\maketitle

\begin{abstract}

We investigate the influence of walls and corners (with Dirichlet
and Neumann boundary conditions) in the evolution of
twodimensional autooscillating fields described by the Complex
ginzburg-Landau equation. Analytical solutions are found, and
arguments provided, to show that Dirichlet walls introduce strong
selection mechanisms for the wave pattern. Corners between walls
provide additional synchronization mechanisms and associated
selection criteria. The numerical results fit well with the
theoretical predictions in the parameter range studied.

\end{abstract}


\begin{twocolumns}
\section{INTRODUCTION.}
\label{sect:intro}

Spatially extended nonlinear dynamical systems display an amazing
variety of behavior including pattern formation,
self-organization, and spatio-temporal chaos
\cite{Cross93,Cross94,Dennin96,Egolf2000}. Much effort has been
devoted to the characterization of the different dynamical regimes
and the transitions between them for model equations such as the
complex Ginzburg-Landau equation (CGLE) \cite{vSaarloos94}. This
is an equation for a complex field $A({\bf x},t)$ that
conveniently rescaled reads
\be
\partial_t A = A + (1+i \alpha) \nabla^2 A - (1+i \beta) |A|^2 A
\label{cgle}
\ee
$\alpha$ and $\beta$ are real parameters. This equation describes
the onset of an oscillatory regime through the Hopf bifurcation of
a homogeneous state, and it is used generally as a model equation
due to the rich variety of its solutions. Binary fluid convection
\cite{Kolodner95}, transversally extended lasers
\cite{Coullet89,SanMiguel95}, chemical turbulence
\cite{Kuramoto74,Kuramoto81}, bluff body wakes \cite{Leweke94},
the motion of bars in the bed rivers \cite{Schielen93}, and other
systems \cite{vanderVaart97} have been described using the CGLE in
a proper parameter range.

The CGLE admits simple plane-wave solutions. However, for most of
the $(\alpha,\beta)$ parameter range, a typical evolution starting
from random initial conditions leads to complex, steady or
evolving, states. An important ingredient in the description of
these dynamical regimes in two-dimensional domains is the
interaction of singular points called {\em defects}. For our
purposes, a defect is just a zero of the complex field $A$, where
there is a singularity in the phase $\varphi$ defined by the
relation $A = |A|\exp (i \varphi)$. There is a topological charge
associated to each defect, $n$, defined by
\be
n = \frac{1}{2 \pi}\oint_\Gamma \vec \nabla \varphi \cdot d \vec r
\label{charge}
\ee
where $\Gamma$ is a closed path around the defect. The topological
nature of the phase singularities implies that $n$ is a positive
or negative integer, and that the total topological charge in the
two-dimensional system is constant, except for the defects flowing
in and out through the boundaries. In the interior of the system,
defects can only be created or annihilated by pairs of opposite
charge. {\em Spiral defects}, i.e. defects around which the lines
of constant phase have a spiral form, are typically formed in the
CGLE (for $\alpha
\ne \beta$). The interaction between these spiral structures has attracted
much attention \cite{Aranson93}. Spiral solutions of a different nature appear
e.g. in excitable media such as the Belousov-Zabotinsky reaction
\cite{Kuramoto84,Winfree87} and electro-hydrodynamic convection
(see e.g., ref \cite{Rehberg89}).

One important source of defects in real systems are the
boundaries. Under some circumstances, walls can introduce defects
into the system increasing the amount of disorder in the dynamics.
In other situations the boundaries play the opposite r\^ole: they
annihilate defects driving the system to a more ordered state. In
general, the interplay between these two behaviors and the proper
dynamics of the bulk can push the system to configurations
different from the ones found under boundary-free conditions
(periodic boundary conditions for instance). However, few studies
have been addressed to the influence of the boundary shapes and
boundary conditions on complex dynamics. The importance of these
effects in the transverse patterns of laser emission, where aspect
ratios are not large, is visible in recent works such as
\cite{Staliunas97,Aranson97}. In addition, average patterns in
Faraday waves and other spatio-temporally chaotic systems have
been observed to be sensible to boundary shape
\cite{Gluckman95,Eguiluz99} and boundaries are able even to induce
spatial chaos in  otherwise non-chaotic systems \cite{Eguiluz99c}.
All those strong influences of boundaries on the dynamics of
extended nonlinear systems \cite{EguiluzPhysA} provides us with
the motivation for a more systematic study of boundary effects on
the CGLE performed in this Paper.

In Ref.~\cite{Eguiluz99a} we performed a first numerical
exploration of these effects, via computer simulations of the CGLE
in circular and rectangular geometries with null Dirichlet
boundary conditions. The results reveal a fundamental r\^ole of
boundaries in selection mechanisms. In particular wave emission
from Dirichlet walls (i.e., walls where $A=0$), and the dominance
of corners as pacemakers for the whole system were important
observed effects. Understanding the origin of such effects is the
main goal of this Paper. To achieve it, we will focus first on the
effect of a single lateral wall, where the complex field is set to
zero, in the selection of the pattern. After this we will study
how the presence of corners (i.e. the intersection of two walls)
restricts the family of solutions found previously. It should be
noted that we use Dirichlet boundary conditions (and in some
places of this Paper also Neumann boundary conditions) as simple
phenomenological conditions to explore deviations with respect to
the more commonly used periodic boundaries. A different issue is
to establish what are the pertinent boundary conditions arising
for the CGLE when it is derived as an amplitude equation in
particular physical contexts (for example in optics, fluids,
etc.). Some results in this last subject can be found in
Ref.~\cite{Roberts92,Martel96}.

In the next Section we review previous numerical results on the
CGLE in several geometries and boundary conditions. In
Section~\ref{sect:unb}, we summarize analytical solutions in
unbounded domains. In Section~\ref{sect:wall}, we present
analytical and numerical results for the CGLE in the presence of a
lateral Dirichlet wall. In Section~\ref{sect:corners}, we extend
our study to the case of corners and in
Section~\ref{sect:conclusions} we finish with our Conclusions.

\section{NUMERICAL OBSERVATIONS.}
\label{sect:numerics}

It is quite evident, and confirmed by our previous study
\cite{Eguiluz99a}, that the effect of boundaries is more
noticeable in the parameter regimes for which large correlation
lengths are present in the system. In strongly chaotic states with
short correlation lengths, the main effects of walls are
restricted to boundary layers close to them. In consequence we
restrict here the presentation of our numerical results to the
region of parameters for which coherent oscillations extend over
nearly the whole system. This happens in most of the Benjamin-Feir
stable region in parameter space, that is, for $1+\alpha
\beta>0$, but also in other regions close to it.
Defects and shocks however disrupt the otherwise ordered plane
waves, and its location and structure are strongly dependent on
boundaries. In Fig.~\ref{fig:ftime}, the CGLE is solved in a
square with null Dirichlet boundary conditions ($A=0$). The
zero-amplitude boundaries facilitates the formation of defects
near the walls. Starting from random initial conditions, defects
are actively created in the early stages of the evolution (See
Fig.~\ref{fig:ftime}). After some time however all the points on
the boundaries synchronize and oscillate in phase so that plane
waves are emitted. Defect formation ceases, and the waves emitted
by the walls push the remaining defects towards the central region
of the domain. There the defects annihilate in pairs of opposite
charge and as a result of this process a bound state is formed by
the surviving set of equal-charge defects. The orientation of the
waves emitted by the boundaries also changes during the evolution.
The synchronized emission of the early stages produces wave
propagation perpendicular to the boundary but in the late states
the wavevector tilts to some emission angle of approximately $45$
degrees. The precise value of this angle depends on both the
parameter values and the geometry of the boundaries. The fact that
this angle is not exactly $45$ degrees is made evident by the
slight mismatch between the waves coming from orthogonal walls.
Finally the system reaches a frozen state of the type displayed in
Fig.~\ref{fig:ffrozen_s}. The term {\em frozen} is used here to
denote that the modulus is a steady solution, although the phase
is time-periodic. More concretely, our frozen configurations are
well-described by $A({\bf x})= f({\bf x})e^{-i\omega t}$, with
$\omega$ real and $f$ a possibly complex function of position.

In the final frozen state, defects are confined in the center of
the domain forming a rigid static chain.
Shock lines appear where waves from different sides of the contour
collide. The strongest shocks are attached perpendicularly to the
walls and the general shock configuration is what one would expect
for small symmetry breaking of the square geometry \cite{maza} The
number of defects depends on the initial condition. Solutions with
no defects are also found (e.g Fig.~\ref{fig:ffrozen_s}(c,d)), and
are called {\em target}-like solutions. This kind of solutions is
not seen in simulations with periodic boundary conditions.

In our simulations in the square geometry with Dirichlet boundary
conditions, the direction of the phase velocity (from the walls or
towards the walls) and the wavenumber depend on the parameter
values in a way which differs from the usual spiral waves in
infinite systems (see Ref.~\cite{Hagan82} and
Section~\ref{sect:unb}).
Thus boundaries are playing an important r\^ole in the selection
of the wave speed and wavenumber.

\begin{figure}
\begin{center}
\epsfig{file=\dirfig/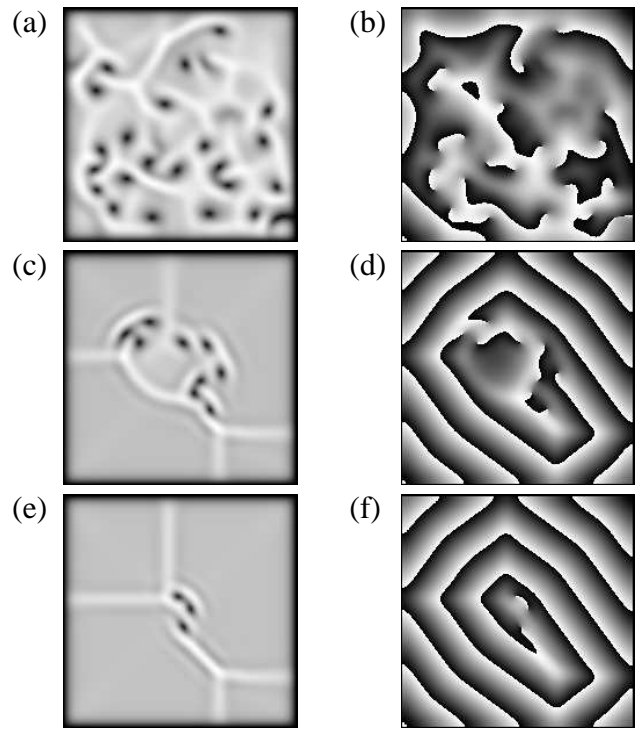,width=0.5\textwidth}\vspace{-3cm}
\caption{\label{fig:ftime} \small{Time evolution of the solution of
the Eq.~(\ref{cgle}) at arbitrary times with parameter values
$\alpha = 2$, $\beta = -0.2$. The initial conditions is random.
When the boundary starts emitting waves, the spiral defects are
pushed to the interior of the domain and annihilate by pairs of
opposite charge. The modulus of the field is plotted in the left
column and the phase in the right, in grey-scale. The final state
(not shown) contains a single defect, as the one in
Fig.~\ref{fig:ffrozen_s}a }}
\end{center}
\end{figure}

In a circular domain (Fig.~\ref{fig:ffrozen_c}), the frozen
structures are either targets (no defects) or a single central
defect. Groups of defects of the same charge can also form bound
states, but instead of freezing they rotate together. This
contrasts with the behavior of the square domains and is
correlated with the absence of shock lines linking the boundaries
to the center in the case of the circular domains. These links are
probably responsible for providing rigidity to the stationary
configuration in the square case. Tiny shock lines associated to
small departures from circularity in the lines of constant phase
can be observed also in the circle but these lines end in the bulk
of the region before reaching the boundaries. On the other hand,
the constant-phase lines reach the boundaries nearly tangentially
in contrast to what we observe in the square. For circular domains
the phase-velocity direction changes with parameters in a way more
similar to the infinite-system spiral.
This is another feature revealing that circular
boundaries introduce less rigidity into the pattern than square
ones. The absence of corners is probably the main qualitative
difference. In fact, corners are observed to act as pacemakers
from which wave emission entrains the whole system
\cite{Eguiluz99a}.

\begin{figure}
\begin{center}
\epsfig{file=\dirfig/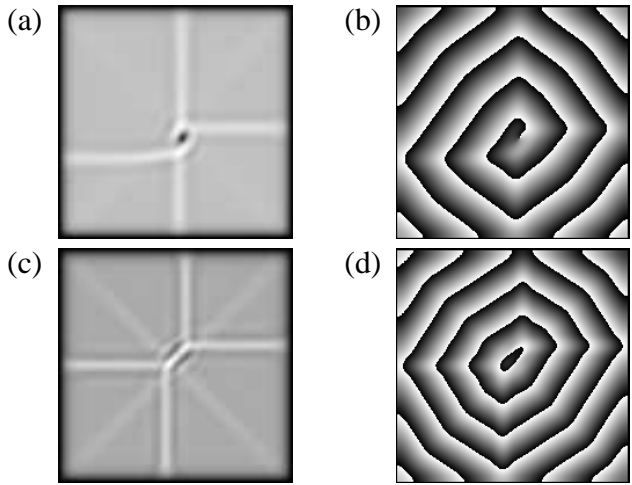,width=0.5\textwidth}
\caption{\label{fig:ffrozen_s} \small{Frozen structures under null
Dirichlet boundary conditions in a square of size $100\times100$.
Parameter values are $\alpha=2$, $\beta=-0.2$ (a,b), and
$\alpha=2$, $\beta=-0.6$ (c,d).  Snapshots of the modulus $|A|$ of
the field are shown in the left column and snapshots of the phase
in the right column. Grey scale runs from black (minimum) to white
(maximum). } }
\end{center}
\end{figure}

\begin{figure}
\begin{center}
\epsfig{file=\dirfig/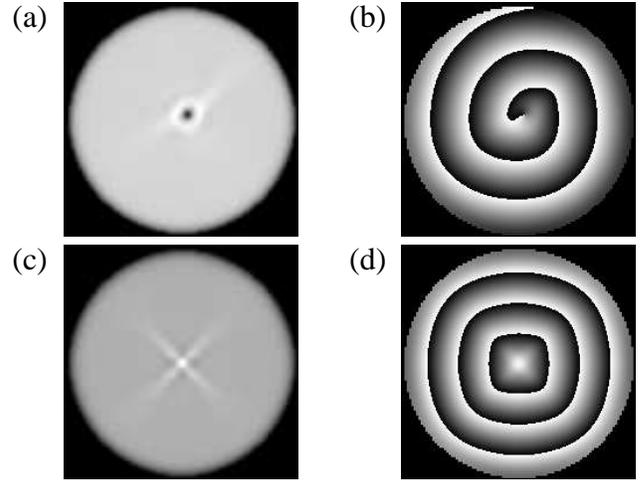,width=0.5\textwidth}
\caption{\label{fig:ffrozen_c} \small{Frozen structures under null
Dirichlet boundary conditions in a circle of diameter $100$ for
parameter values $\alpha=2$, $\beta=-0.2$ (a,b), and $\alpha=2$,
$\beta=-0.6$ (c,d).  Snapshots of the modulus $|A|$ are shown in
the left column and the phase is shown in the right column. Gray
scale as in Fig.~\ref{fig:ftime}. } }
\end{center}
\end{figure}

The stadium shape (Fig.~\ref{fig:ffrozen_d}) mixes features of the
two geometries previously studied: it has both straight and
circular borders. In this case, the curves of constant phase
arrange themselves to combine the two behaviors described above.
On the one hand the lines meet the straight portions of the border
of the stadium with some characteristic angle, as it happens in
square domains. However, these lines bend to become nearly tangent
to the semi-circles in the places where they meet with these
portions of the boundaries. A typical frozen solution displays a
shock line connecting the centers of the circular portions of the
domain. This shock line usually contains defects and their
dynamics in this stage is much slower than the annihilation that
occurs in the bulk of a domain without the presence of shocks. It
is also possible to find defect-free target solutions as in the
case of the circle, and the behavior of the phase velocity is also
similar in the way its direction can be changed by modifying the
parameters.

\begin{figure}[t]
\begin{center}
\epsfig{file=\dirfig/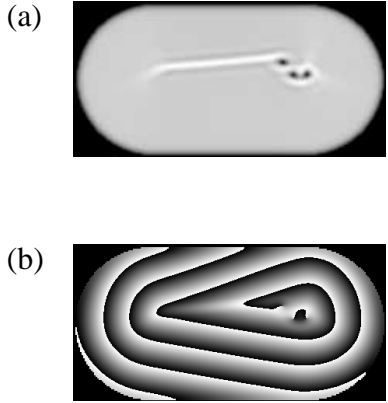,width=0.5\textwidth}
\caption{\label{fig:ffrozen_d} \small{Frozen structures under null
Dirichlet boundary conditions in domain with stadium shape of axis
$200 \times 100$ for parameter values $\alpha=2$, $\beta=-0.2$.
Snapshots of the modulus $|A|$ are shown in the left column and
the phase is shown in the right column. Gray scale as in
Fig.~\ref{fig:ftime}. } }
\end{center}
\end{figure}

To summarize, Dirichlet boundary conditions play a double r\^ole.
On one hand, the walls naturally behave as sources (or sinks) of
defects. On the other hand, a wall with null Dirichlet conditions
shows a tendency to emit plane waves that will coherently fill the
whole system. The interplay between these two properties of the
boundaries gives rise to interesting behavior. In the case of
frozen states, the character of the walls as wave emitters
dominates. The intersection of two walls (a corner) is observed
also to emit waves, and the whole system becomes synchronized to
this corner emission. In circular domains, on the other hand,
there are no corners and wave selection is definitively dominated
by the internal spirals. Another aspect of the dynamical dominance
of the walls in the square case is that defects form a chain which
is anchored to the boundaries by a set of shock lines; in a
circle, on the contrary, the asymptotic state is usually a bound
state disconnected from the boundaries.

Gaining some understanding of aspects of our numerical
observations is the goal of the next Sections.

\section{SOLUTIONS IN UNBOUNDED DOMAINS.}
\label{sect:unb}

In this Section we review some of the analytical solutions of the
CGLE in unbounded domains. First we start with plane waves,
continue with one-dimensional holes and finish with
two-dimensional spirals.

The CGLE possesses, among many other solutions, a family of plane
wave solutions and solutions containing phase-singular points. The
plane-wave continuous family is parameterized by the corresponding
wave number ${\bf k}$. The form of the solutions is $A=R \exp [i
({\bf k}\cdot {\bf x} -
\omega t)]$, where $R= \sqrt{1-k^2}$, $\omega(k) =
\beta - k^2 (\beta- \alpha)$, and $k=|{\bf k}|$. The  limit of
stability of plane waves is known as the Benjamin-Feir line and is
given by the curve $1+\alpha \beta =0$; if this quantity is
positive, some stable plane wave exists \cite{Pere}; if $1+\alpha
\beta < 0$ all plane waves are unstable. The limit is given by the
stability of the plane wave with $k=0$. Stability analysis gives
that plane waves possessing wave number $k$ in the range
$[-k_c,k_c]$, where $k_c = {\sqrt \frac{1 +
\alpha \beta}{3 +  \alpha \beta + 2 \beta^2}}$, are stable. The
instability is with respect to long-wavelength disturbances whose
wave vectors are parallel to ${\bf k}$ (Eckhaus instability)
\cite{Pere}. It will be useful for the  future discussion to have
an expression for the {\em phase velocity} of the waves, and of
the {\em group velocity} of small perturbations on such waves,
${\bf v}_{ph}$ and ${\bf v}_{gr}$ respectively
\ba
{\bf v}_{ph} &=&  \frac{\omega(k)}{k}{\bf \hat k}~,
\label{vph_pw}
\\ {\bf v}_{gr} &=& -2 k (\beta - \alpha){\bf \hat k}~.
\ea
${\bf \hat k}$ is the unit vector in the direction of {\bf k}. The
expression for the group velocity \cite{Montagne97} turns out to
be equivalent to the linearly-looking expression ${\bf v}_{gr} =
\nabla_{\bf k}\omega(k)$, even though $\omega(k)$ is the
dispersion relation of nonlinear waves.

In addition to simple waves, the one-dimensional CGLE possess a
one-parameter family of solutions for which the amplitude displays
a region of local depression. Their analytic form was determined
by Nozaki and Bekki \cite{Nozaki84}, and they are therefore also
referred to as Nozaki-Bekki solutions or {\it holes}. One member
the family is characterized by the value of $A$ being zero at a
point, called the {\em core} of the hole, and asymptotically
behaving, at both sides of the core, as a plane wave of wave
number $k$.  It is worth noting this one-dimensional hole solution
was also obtained by Hagan \cite{Hagan82} as a sub-product of his
calculations for two-dimensional spirals. At variance with the
other members of the Nozaki-Bekki family, this hole does does not
travel into the system and thus it will be denoted as the {\sl
standing hole}. Its analytical expression (choosing the origin of
coordinates at the hole core) can be written as
\be
W_H(x,t) = \sqrt{1-k^2} \tanh (p x) \exp [i (\psi (x) - \omega
t)]~,
\label{static1dhole}\ee
where $\psi$ is a function of $x$ satisfying
\be
d \psi / dx = k \tanh (p x)~,
\ee
(i.e. $\psi=\psi_0+(k/p)\log\cosh (px)$, with $\psi_0$ an
arbitrary reference phase) and $\omega$, $k$, and $p$ are related
according to
\ba
\omega &=& \beta - k^2 (\beta - \alpha)~, \\ k   &=& \frac{2
p^2 -1}{3 p \alpha}~,
\\ \{4 ( \beta - \alpha) + 18 \alpha (1+&\alpha^2&)\} p^4
\nonumber\\ - \{4 ( \beta - \alpha) + 9&\alpha&(1 + \alpha
\beta)\}p^2 + \beta -\alpha = 0 ~. \ea

If $\alpha = 0$ we can easily get $p=1/\sqrt{2}$ and
\ba
\omega &=& \beta (1- k^2)~, \\
p   &=& 1/\sqrt{2}~, \\
\beta &=& -\frac{3 k}{\sqrt 2 (1-k^2)}~.
\label{h1d}
\ea
Thus $\beta$ and $k$ have opposite sign ($\beta k< 0$), when
$\alpha = 0$. For any value of $\alpha$ and $\beta$, the existence
of a defect-like solution fixes the value of the asymptotic wave
number $k$ and accordingly that of $\omega$. We mention here that
for configurations of the {\sl frozen} type, the solutions with
$\alpha$ arbitrary can be obtained from the ones with $\alpha=0$
by means of a change of variables. This fact, which frequently
simplifies analysis, is detailed in the Appendix.

The phase and group velocity far from the core for the
one-dimensional standing hole with $\alpha = 0$ are
\ba
v_{ph} &=& \frac{\beta (1-k^2)}{k} = -\frac{3}{\sqrt 2} < 0~, \\
v_{gr} &=& - 2 k \beta > 0 ~.
\ea
Thus the propagation of the phase is towards the core of the
defect independently of the value of $\beta$. However, the group
velocity is directed outwards from the core of the defect. Thus
small perturbations to this solution are expelled away from the
core. The case of arbitrary $\alpha$ can also be solved
numerically. Given the parameters $(\alpha,\beta)$, the line where
the phase velocity is zero can be found and it is plotted in
Fig.~\ref{fig:fvph}. The group velocity turns out to be always
positive (i.e. outwards from the core) for the standing hole
solutions.

The two-dimensional spiral wave solutions of the CGLE are
solutions winding around a defect core (i.e. a phase singularity).
In polar coordinates $(r,\theta)$ around the core, they have the
analytical form \cite{Hagan82}:
\be
D(r,\theta,t) = R(r) \exp (i (\theta + S(r) - \omega t))~.
\label{spiral}
\ee
This solution represents a phase pattern rotating steadily around
$r=0$ with frequency $\omega$ (and frozen modulus). The amplitude
$R$ is a monotonically increasing function of $r$, proportional to
$r$ near the origin, and asymptotically approaching some value
$R_\infty < 1$ for large $r$. The function $S$ behaves smoothly in
the neighborhood of the origin, taking the form $S \sim S_0 + S_1
r^2$. Far from the origin $S$ becomes proportional to $r$,
behaving as $S
\sim k r$. In this way, in the distant region, the isophase lines
approach the form of Archimedian spirals, converging to plane
waves with wave number $k$. Thus $R_\infty = \sqrt{1-k^2}$ and
$\omega =
\beta - k^2 (\beta-\alpha)$. The charge of solutions of the form
Eq.~(\ref{spiral}) is, according to eq.~(\ref{charge}), equal to
$+1$. There exists also a negatively charged spiral, with the form
Eq.~(\ref{spiral}) but with $\theta$ replaced by $-\theta$. In
spiral waves, wave motion is induced in such a manner as to cause
the global synchronization of the medium by the defect.

It is important to notice that, in both one (the standing hole)
and two dimensions (the spiral solution), imposing the
requirements of zero field at the core, and plane wave behavior
far from the core, the value of $k$ gets fixed. Thus fixing the
parameter values $(\alpha,
\beta)$, the spiral structure (and the standing hole) is unique
(except for an arbitrary change in the location of the core). The
precise way in which wavenumber, frequency, phase or group
velocities depend on parameter values $(\alpha,\beta)$ can be
found for example in Ref.~\cite{Hagan82}.

\begin{figure}[t]
\centerline{\epsfig{file=\dirfig/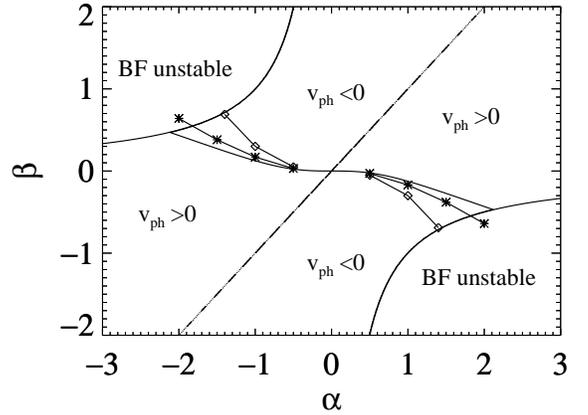,width=0.5\textwidth}}
\caption{\label{fig:fvph}
\small{Parameter space of the CGLE. Different regions are separated
by solid lines: BF unstable regime where there are no stable plane
wave solutions in the infinite system; regions where the {\em
phase velocity} $v_{ph}$ is positive or negative are also shown,
and separated by  additional solid lines for the case in which a
single Dirichlet wall is present in the system. Star line
corresponds to zero phase velocity for emission from a corner
between two Dirichlet walls spanning an angle of 135 degrees;
diamond line corresponds to a 90-degrees corner.}}
\end{figure}

\section{SOLUTIONS WITH A SINGLE WALL.}
\label{sect:wall}

As a first step to understanding the solutions of the CGLE in
bounded domains, we study in this Section solutions in the
presence of a single wall where the value of the complex field $A$
is set to zero. We observe numerically that, starting from random
initial conditions in a bounded domain with a single Dirichlet
wall, frozen solutions are reached asymptotically (see
Fig.~\ref{fig:fwall}). In our numerical implementation, the
Dirichlet wall ($A=0$) is the left one, Neumann boundary
conditions (zero normal derivative of $A$) are applied to the
right wall, and the upper and lower limits of the domain are
identified via periodic boundary conditions. Initially some
(spiral) defects are formed. Typically, the Dirichlet wall starts
to emit plane waves that push the defects towards the opposite
boundary until they are all expelled or annihilated. The
stationary solution is the two-dimensional extension of the
one-dimensional standing hole solution described in
Section~\ref{sect:unb} (that is a continuous line of holes with
their cores on the wall: $W_H (x,y,t) = W_H(x,t)$).

We can investigate the possibility of more complex solutions in
which the amplitude is independent of the $y$-direction and takes
the form of a hole solution in one dimension, but with a phase
that depends explicitly on both coordinates. We study first the
case of $\alpha = 0$ to come back later to the general case.

We look for solutions of the form
\be
W_W(x,y,t)= \sqrt{1- k^2} \tanh ( p x ) \exp[i(\psi(x,y) -
\omega t)]
\label{tiltedhole}
\ee
with $\omega = \beta (1-k^2)$ and $k^2=k_x^2+k_y^2$. Assuming the
form $\psi(x,y)=
\psi(x) + \psi(y)$, we get the relations
\ba
\partial_x \psi (x,y) &=& k_x \tanh(p x) \label{dxtheta}\\
\partial_y \psi (x,y) &=& k_y     \label{dytheta} \\
2 p^2          &=& 1 - k_y^2  \label{ec11}    \\
3 k_x p         &=& \beta (1-k^2) \label{ec12}
\ea
and substitution of Eq.~(\ref{ec11}) in Eq.~(\ref{ec12}) gives
\be
\frac{3 k_x \sqrt{1- k_y^2}}{\sqrt{2}} = - \beta (1- k^2)~.
\label{swdisp}
\ee
Note that if $k_y = 0$ we recover the expression for the
one-dimensional standing hole solution (in particular we recover
Eq.~(\ref{h1d})).

We can perform a similar calculation for the general case of
parameters $\alpha$ and $\beta$. For a solution of the form
Eq.~(\ref{tiltedhole}), Eqs.~(\ref{dxtheta}) and (\ref{dytheta})
remain valid, and $\omega$, $k$, and $p$ are related according to
\ba
\omega &=& \beta - k^2 (\beta - \alpha)~, \label{omegaHole}\\
k_x  &=& - \frac{2 p^2 + k_y^2 -1}{3 p \alpha}~,  \label{kxHole}\\
0 &=& - 3 p k_x + \alpha (2 p^2 - k_x^2) -\beta (1 - k^2) ~.
\label{kpkxHole}\ea

In contrast with the selection mechanism for the standing hole or
the spiral solutions, the presence of the wall does not select a
unique wave vector but a one-parameter family of solutions
parametrized by either $k_x$ or $k_y$ arises instead from Eqs.~
(\ref{dxtheta})-(\ref{swdisp}) or
(\ref{omegaHole})-(\ref{kpkxHole}) for given values of $\alpha$
and $\beta$. Different solutions in the family differ in the
direction and magnitude of the wavevector $\bf{k}$. Different
wavevectors change the angle of intersection between the lines of
constant phase and the wall. Figures~\ref{fig:fwall}(e-f) are the
final state in a numerical simulation in which the initial
condition was close to (\ref{tiltedhole}) with $\bf k$ oblique
with respect the wall. The displayed state is identical (far
enough from the Neumann wall) to Eq.~(\ref{tiltedhole}) with
Eqs.~(\ref{omegaHole})-(\ref{kpkxHole}) thus numerically proving
the stability of this solution. Different orientations of $\bf k$
can be tested in the same way. However, if starting with random
initial conditions we typically find solutions corresponding to
the case $k_y =0$ (Fig.~\ref{fig:fwall}c-d) that is the simplest
two-dimensional extension ($k_y=0$) of the standing hole. The
prevalence of the $k_y=0$ solution when starting from random
initial conditions (of small amplitude) can be understood by
analyzing the linear stability of the state $A=0$ with Dirichlet
conditions in a single wall limiting a semi-infinite domain. The
linear eigenmodes are of the form $a(x)\exp(k_y y)$, and it is
easy to see that the fastest growing one has $k_y=0$ so that it
will overcome the other ones at long times before nonlinear
saturation.

We note that, although solutions (\ref{tiltedhole}) represent
emission at an angle with the wall, the analytic expression
predicts a small boundary layer (of size $p^{-1}$) in which the
wavenumber leaves its asymptotic orientation to become parallel to
the wall, so that isophase lines arrive perpendicular to the
boundary. This is observed in the numerical solutions (see for
example Fig.~\ref{fig:fwall}f, and also the rectilinear walls of
Figs.~\ref{fig:ftime}, \ref{fig:ffrozen_s}, and
\ref{fig:ffrozen_d}) thus nicely confirming the relevance of the
analytical solutions to the observed configurations. The
analytical expressions describing wave emission with phase lines
parallel or at small angle with the walls also provide approximate
descriptions of the wave emission in the circular geometry. In the
square, however, conflict between the orientation emitted by
neighboring walls occurs, and the exact expression
(\ref{tiltedhole}) is appropriate only near each wall. The
conflict between neighboring walls is resolved at long times by
emission from the corner, as will be seen in the next Section.

Another important kind of solutions with a single wall is the one
that appear with Neumann boundary conditions. The solutions
observed close to the right wall in Fig.~\ref{fig:fwall}(c-f) are
of this type. These solutions have been already analyzed in the
literature, specially in the context of interactions between
spirals. The reason is that a Neumann wall acts as a reflecting
boundary, so that having a wave impinging into the boundary is
equivalent to the interaction between two sources of waves located
symmetrically with respect to the wall \cite{Aranson93}. Despite the interest
of such solutions, no exact analytical expression has been found
for them. Analytical solutions have been obtained
however\cite{phase} by solving the phase equation that
approximates the phase dynamics for slow amplitude variations. In
agreement with the numerical observations, the solution presents a
maximum modulus at the wall (a shock) and the isophase lines,
straight in the far field, deform when entering a boundary layer
close to the wall to arrive parallel (for normal incidence) or
perpendicular (for tilted far-field incidence) to the wall. We
will see in the following that these `tilted Dirichlet waves' are
of relevance when corners are present.

It is clear on physical grounds, and confirmed by the analytical
expressions from the phase approximation, that the wall can act as
a sink of waves of arbitrary far-field orientation and wavelength
(the maximum modulus at the shock will adapt accordingly). Neumann
waves constitute, then, a biparametric family for fixed
$(\alpha,\beta)$.

\begin{figure}
\centerline{\epsfig{file=\dirfig/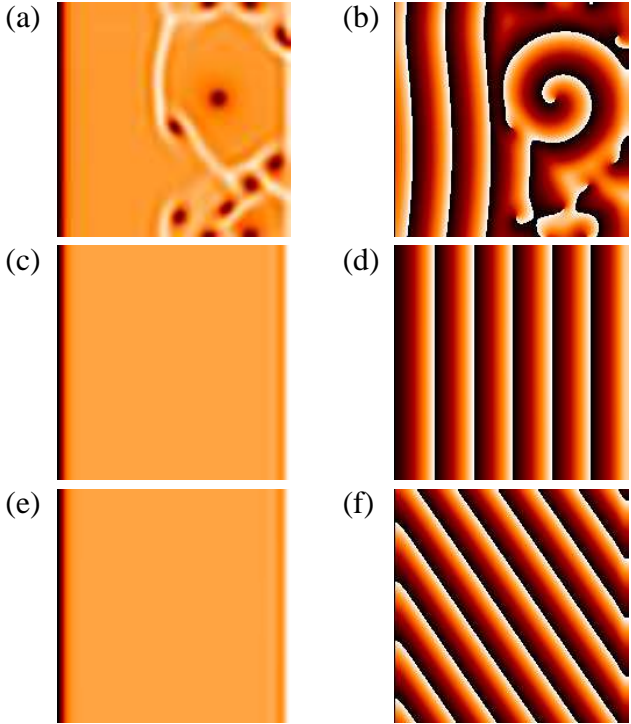,width=0.5\textwidth}}
\vspace{-3cm}
\caption{\label{fig:fwall}
\small{Modulus (left column) and phase (right column) of the solution of
Eq.~(\ref{cgle}) for $\alpha = 2$ and $\beta = -0.2$ with
Dirichlet boundary conditions for the left boundary, Neumann for
the right one, and periodic in the horizontal ones. (a,b):
Early-time state starting from random initial conditions of small
amplitude.(c-d): The final asymptotic state. The lines of constant
phase travel to the right. Notice that although there is a
developed spiral defect it disappears through the Neumann boundary
at long times. (e,f): The long-time asymptotic state from an
initial condition consisting of distorted plane wave with
wavevector oblique to the boundaries. A solution of the form
Eq.~(\ref{tiltedhole}), with wavevector close to the initial one,
is finally reached.}}
\end{figure}

\section{SOLUTIONS IN PRESENCE OF CORNERS.}
\label{sect:corners}

We now pay attention to the effects induced by the presence of
corners, i.e. how the solutions adapt to the emission of two
semi-infinite lines. In subsection~\ref{sect:velocity} we will
show that the phase velocity not only depends on the parameters of
the CGLE $\alpha$ and $\beta$, but also on the angle $\phi$
between the walls of the boundary. In subsection~\ref{sect:V}, we
will present solutions of the phase equation representing wave
collision; they are usually called V-solutions. These solutions
should be matched with the boundaries, which provide selection
mechanisms for the wave pattern.

\subsection{Phase velocity dependence on the angle at the boundary.}
\label{sect:velocity}

We have performed numerical simulations of the CGLE in the domain
depicted in Fig.~\ref{fig:fbc}, where one of the walls is a broken
line with a corner of a definite angle $\phi$. The boundary
conditions are the following: for the right, upper, and bottom
walls, Neumann boundary conditions (null normal derivative). For
the left boundary (where the corner is present), null Dirichlet
conditions. This left boundary is a line that is broken forming a
variable angle $\phi$. If this angle is $180^\circ$, there exist
the two-dimensional extension of the standing hole described in
the previous Section. As the angle decreases the wave is not
longer plane, and the phase velocity adapts to the new geometry.
The wave fronts may become just slightly distorted from straight
lines (as in Fig.~\ref{fig:fcorner}) or display a kink (similar to
the situation in Fig.~\ref{fig:ffrozen_s}) depending on $\alpha$,
$\beta$, and $\phi$. In any case the kink is never too strong and
departures from straight wavefronts never large. Changing
parameters the phase velocity may vanish. The locus in parameter
space where this happens is a 2d surface in the
$(\alpha,\beta,\phi)$ space. Projections in the $(\alpha,\beta)$
plane for $\phi=180^\circ$ (obtained from
Eq.~(\ref{static1dhole})) and $\phi = 135^\circ$ (from numerical
simulation) are plotted in Fig.~\ref{fig:fvph}. $\phi=90^\circ$
corresponds to a square and is also plotted in
Fig.~\ref{fig:fvph}. We do not see differences between squares
with two or four Dirichlet walls.

\begin{figure}
\begin{center}
\epsfig{file=\dirfig/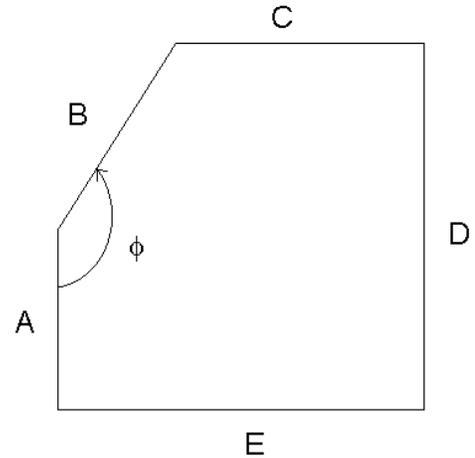,width=0.4\textwidth}
\caption{\label{fig:fbc}
\small{Domain and boundary conditions. In A ad B, null Dirichlet boundary
conditions; C, D and E, null Neumann boundary conditions}}
\end{center}
\end{figure}

Summarizing, for frozen structures, the presence of Dirichlet
walls establish a selection mechanism different from the
associated to the presence of a spiral core in an infinite domain
(Section~\ref{sect:unb}). When the Dirichlet wall is broken, it is
seen in the earliest stages of wave-pattern development that
emission with isophases parallel to the walls is initiated, but
collision between the waves from different walls arises and a
distinct final state, with wavenumber, phase velocity, etc. fixed
by $(\alpha,
\beta, \phi)$ is reached. We now investigate how this may happen.

\subsection{V-solutions of the phase equation and pattern selection.}
\label{sect:V}

For unbounded domains and for small amplitude modulations, a phase
description of the complex field $A$ can be made. The approximate
equation for the phase is \cite{Kuramoto84}
\be
\dot \varphi = \omega_0 + b_0 \nabla^2 \varphi + b_1 (\nabla \varphi )^2 \\
\label{phaseEq}
\ee
where $b_0 = 1 + \alpha \beta$, $b_1 = (\beta - \alpha)$, and
$\omega_0 = - \beta$.

We look for solutions $\varphi = \varphi (x,y,t)$ representing
phase waves with non-straight isophase lines. This is what is
observed when Dirichlet waves from different waves interact
(Figs.~\ref{fig:ffrozen_s},\ref{fig:fcorner}). Analytic
expressions of this type are known for the phase equation: the
V-solutions \cite{Kuramoto84}. We impose different but symmetric
wavevectors far from a shock occurring at $x=y$, that is ${\bf
\nabla}\varphi\rightarrow (k_1,k_2)$ if $x \ll y$, and ${\bf
\nabla}\varphi\rightarrow (k_2,k_1)$ si $y \gg x$, thus getting
the family of solutions:
\ba
\varphi(x,y,t) &=& \left(\omega_0 +
b_1(k_1^2+k_2^2)\right)t + \frac{k_1+k_2}{2} (x+y)  \nonumber   \\
&+& \frac{b_0}{b_1} \log\left[\cosh\left(
\frac{b_1}{b_0} \frac{k_1-k_2}{2} (x-y)
\right)\right]   \ ~.
\label{theVsolution}
\ea

The spatial dependence of this solution can be separated in terms
of the variables $u=x+y$ and $v=x-y$, and thus the phase equation
is also separable in $u$ and $v$. The change from $(x,y)
\to (u,v)$ is a rotation bringing the shock line to one of the axes.
After inspection of the derivatives normal to the shock, we see
that half of these V-solutions can be interpreted as tilted waves
approaching a Neumann wall at the shock position, being the other
half just a specular image. As for Neumann tilted waves, we have a
biparametric family, parametrized by $k_1$ and $k_2$.

As solutions of the phase equation, the V-solutions are strictly
valid only far from the boundaries, where the modulus of the field
remains nearly constant. Matching to solutions of the form of
Eq.~(\ref{tiltedhole}) should be performed close to Dirichlet
boundaries. We know (Sect. ~\ref{sect:wall}) that for this type of
boundaries, the two components of the far-field wavevector are not
independent (Eqs.~(\ref{omegaHole})-(\ref{kpkxHole})). This
introduces a relationship between $k_1$ and $k_2$ in
(\ref{theVsolution}). If the shock line $x=y$ bisects the angle
$\phi$ between two Dirichlet walls, no additional constraints
appear from matching to the other boundary. Thus, one of the
parameters in the V-solution, which can be taken as the angle
between the two waves, is still undetermined.
From the numerical simulations, it appears that this angle becomes
determined when the medium is synchronized by the waves coming
from the corner between the two walls. We do not have a rigorous
argument to demonstrate that this is the case, but the following
heuristic argument is a step towards such a demonstration: Close
to the walls a phase description is not longer valid, and the
modulus approaches zero. The solution is of the frozen type, which
we write as $A(x,y,t)=R(x,y)\exp(i (\psi (x,y) - \omega t))$ with
real $R$, $\psi$, and $\omega$. Since this solution should become
(\ref{theVsolution}) far from the walls, we immediately find
\be
\omega=\omega_0+b_1(k_1^2+k_2^2)=-\beta+(\beta-\alpha)(k_1^2+k_2^2)
\label{matchingw}
\ee
Sufficiently close to the corner at $x=y=0$, the modulus $R$ is
small. Writing for it and for $\psi$ a Taylor expansion, imposing
symmetry across the $x=y$ line, and substituting into the CGLE
(\ref{cgle}), we easily find at the lowest order in distance to
the corner the following behavior:
\ba
R(x,y) &\approx& B xy \\
\psi(x,y) &\approx& \psi_0 - \frac{\omega}{12}(x^2 + y^2) \ ~.
\ea
Close to the walls the local wavevector is ${\bf q} = \nabla \psi
= - \frac{\omega }{6} (x,y)$, so that in the diagonal ${\bf q}_d =
- \frac{\omega }{6} (x,x)$ with modulus $q_d =
\frac{x \omega}{\sqrt 18}$. Far from the corner this wavenumber should
match the one obtained from the V-solution ${\bf k}_d = (k_1 +
k_2)/\sqrt{2}$. An approximate way of doing this is imposing that
both wavenumbers become equal at some distance $x \approx a$ from
the corner:
\be
\frac{\omega}{\sqrt{18}} a \approx \frac{k_1 + k_2}{\sqrt{2}}
\label{kmatching}
\ee
$a$ is an unknown constant of the order of the boundary layer size
($ p^{-1}$). For given parameter values $\alpha$ and $\beta$ this
expression gives an extra relationship between $k_1$ and $k_2$ or,
equivalently, between the modulus $k$ and the angle of emission
from the walls. This, and Eqs.~(\ref{omegaHole}-\ref{kpkxHole}),
completely fixes the solution in the presence of a corner.

Of course, precise numerical values can not be obtained since $a$
is unknown, but the previous heuristic argument was intended only
to illustrate how the presence of the corner resolves the conflict
between the neighboring waves, and fixes the wave pattern as
numerically observed. For situations such as the ones depicted in
Fig. ~\ref{fig:fcorner} for which the wavefronts remain relatively
straight, we have $k_1 \approx k_2$, which can be used as a
substitute of (\ref{kmatching}) to fix the pattern. In fact this
is never a bad approximation. For example, from a $90^\circ$
corner, straight and symmetric wavefronts indicates wave emission
at $45^\circ$ from each wall. We have checked that this is in fact
equivalent to Eq.~(\ref{kmatching}) with $a^2 = 8$. Since this
value of $a$ is within the boundary layer range, both approaches
((\ref{kmatching}) and $k_1 \approx k_2$) are mutually consistent
as can be thought as two different approximations to the same fact
that the corner fixes the wavenumber. Assuming $k_1 = k_2$, we
have plotted in Fig.~\ref{fig:fselec} a comparison between the
results from the numerical simulations and the analytical
predictions. The agreement is good and confirms the relevance of
the walls and corners into the wave selection process.

\begin{figure}
\begin{center}
\epsfig{file=\dirfig/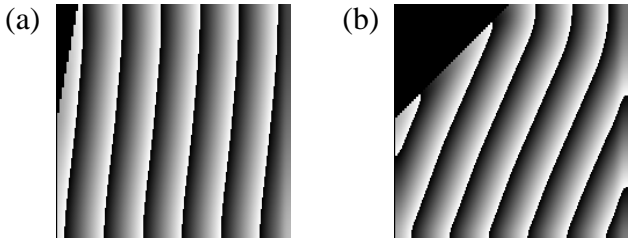,width=0.5\textwidth}
\vspace{-9.5cm} \caption{\label{fig:fcorner} \small{Phase of the
solution of Eq.~(\ref{cgle}) in grey-scale, for parameter values
$\alpha =2$, $\beta = -0.2$. In (a) the angle $\phi = \pi/2
+\tan^{-1} (1/5)$; (b) $\phi = 3 \pi / 4$. }}
\end{center}
\end{figure}

\section{CONCLUSIONS.}
\label{sect:conclusions}

In this Paper we have presented numerical results on the influence
of boundaries in the wave pattern selection of a selfoscillatory
medium, in parameter regimes in which frozen structures are
reached. Analytical solutions in the presence of walls and corners
have been presented and shown to be relevant to the numerically
observed configurations. The dominance of Dirichlet walls, the
relative passive r\^{o}l of Neumann walls, and the synchronization
properties of corners, are possibly generic features which should
be found in other selfoscillatory model systems. Extrapolation to
real experimental oscillatory media should be made with care,
however, since determining the correct boundary conditions
applying to the amplitude equation associated to a particular
medium is a subtle task \cite{Roberts92,Martel96}.

\begin{figure}
\begin{center}
\epsfig{file=\dirfig/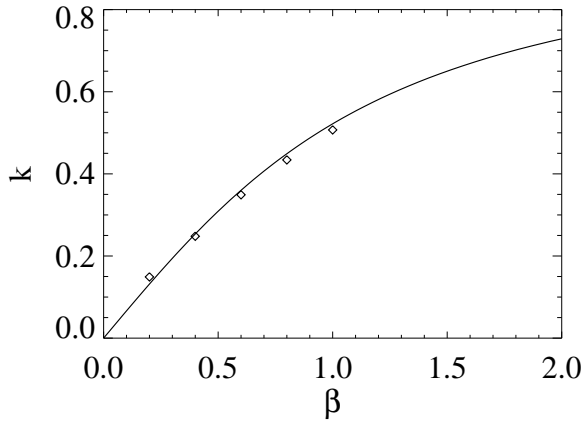,width=0.5\textwidth}
\caption{\label{fig:fselec}
\small{Modulus of ${\bf k}$ versus parameter $\beta$ (for a
square with Dirichlet walls, and $\alpha=0$) obtained from our
theoretical arguments (Eqs.~(\ref{omegaHole})-(\ref{kpkxHole}) and
$k_x=k_y$, solid line) and direct numerical simulation
(diamonds).}}
\end{center}
\end{figure}

Note: In
the following address we have made available a web page containing
simulations of the CGLE in different geometries related to this
paper: {\tt
http://www.imedea.uib.es/$\sim$victor/Video/videos$\_$cgle.html}

\section*{ACKNOWLEDGMENTS.}

Financial support from DGES, Spain, projects PB94-1167, and
PB97-0141-C01-01, is greatly acknowledged. VME aknowledges financial support 
from the Danish Natural Science Research Council.

\section*{Appendix A}

The equation resulting from restricting the CGLE Eq.~(\ref{cgle})
to frozen solutions of the form $A({\bf x})= f({\bf x})e^{-i\omega
t}$, with $\omega$ a real frequency and $f$ a possibly complex
function of the position, admits a change of variables
\cite{Hagan82} that transforms the case with parameters $(\alpha,
\beta)$ into the case with parameters $(0,\beta')$. The transformation is
\ba
\beta '  &=& \frac{\beta -\alpha}{1+\alpha \beta}   \\
\omega ' &=& \frac{\omega - \alpha}{1 + \omega \alpha}  \\
{\bf x}'      &=&   {\bf x} \sqrt{
\frac{1+\alpha\omega}{1+\alpha^2} } =
{{\bf x} \over \sqrt{1-\alpha\omega'}}\\
 f'     &=& f \sqrt
\frac{1 + \alpha \beta}{1+ \omega \alpha}~,
\ea
Obtaining a frozen solution (i.e., a function $f'({\bf x}')$ and
an associated frequency $\omega'$) at parameters $\alpha=0$ and
$\beta'$ thus allows finding corresponding solutions $(f,\omega)$
at arbitrary values of $\alpha$ and the corresponding
$\beta=\frac{\beta'+\alpha}{1-\alpha\beta'}$. This useful
relationship has been used along this Paper to generate solutions
at arbitrary parameters from easier solutions at $\alpha=0$. Note
that if $f'$ contains a factor of the form $e^{i{\bf k}'\cdot{\bf
x}'}$, then $f$ will have a factor of the form $e^{i{\bf
k}\cdot{\bf x}}$, with ${\bf k}={\bf k}'/\sqrt{1-\alpha\omega'}$.


\end{twocolumns}

\begin{thebibliography}{10}
\bibitem{Pere} The stability mentioned here is just with respect
to perturbations of long wavelenght. It is known that additional
instabilities, at finite wavenumber, may appear even if
$1+\alpha\beta>0$; see for example R. Montagne, P. Colet, Phys.
Rev. E {\bf 56} 4017 (1997); these instabilities are not of
relevance for the present work.
\end{thebibliography}
\end{document}